\documentclass[aps,prl,twocolumn,groupedaddress,showpacs]{revtex4}

\usepackage{graphicx}

\begin{document}

\title{Optical Weak Link between Two Spatially Separate Bose-Einstein Condensates}

\author{Y. Shin}
\author{G.-B. Jo}
\author{M. Saba}
\author{T.~A. Pasquini}
\author{W. Ketterle}
\author{D.~E. Pritchard}

\homepage[URL: ]{http://cua.mit.edu/ketterle_group/}

\affiliation{Department of Physics, MIT-Harvard Center for
Ultracold Atoms, and Research Laboratory of Electronics,
Massachusetts Institute of Technology, Cambridge, Massachusetts,
02139}

\date{\today}

\begin{abstract}
Two spatially separate Bose-Einstein condensates were prepared in
an optical double-well potential. A bidirectional coupling
between the two condensates was established by two pairs of Bragg
beams which continuously outcoupled atoms in opposite directions.
The atomic currents induced by the optical coupling depend on the
relative phase of the two condensates and on an additional
controllable coupling phase. This was observed through symmetric
and antisymmetric correlations between the two outcoupled atom
fluxes. A Josephson optical coupling of two condensates in a ring
geometry is proposed. The continuous outcoupling method was used
to monitor slow relative motions of two elongated condensates and
characterize the trapping potential.
\end{abstract}


\maketitle

Josephson effects~\cite{Josephson62} are quantum phenomena in
which the current between two weakly coupled, macroscopic quantum
systems depends on the relative phase of the two systems. These
effects are direct evidence for the existence of the phase of a
macroscopic quantum system~\cite{Anderson64} and observed in
quantum systems such as superconductors~\cite{AR63}, superfluid
$^3$He~\cite{PLB97}, and Bose condensed gases~\cite{CBF01,AGF05}.
Josephson coupling between two systems is typically established
via tunneling through a separating potential barrier or via an
external driving field as in the internal Josephson
effect~\cite{WWC99, OS99}. Both couplings require spatial overlap
of the two systems due to the intrinsic locality of the coupling
interactions.

The concept of Josephson coupling can be extended to include two
\emph{spatially separate} quantum systems by using intermediate
coupling systems. If the phase relations among these systems are
preserved and thus the net particle exchange is phase-sensitive,
the two spatially separate systems might be regarded as being
effectively Josephson-coupled via the intermediate systems.
Furthermore, the phase of the coupling may be actively controlled
by adjusting the coupling states of the intermediate systems. This
idea has been theoretically introduced in the context of relative
phase measurement~\cite{IK97}.

\begin{figure}
\begin{center}
\includegraphics{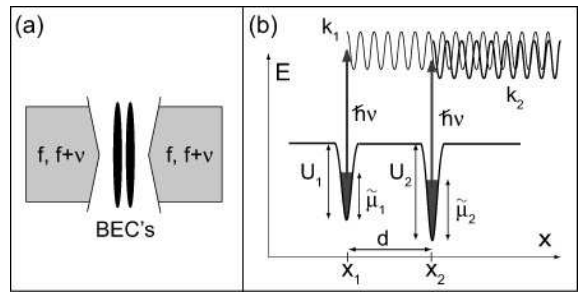}
\caption{Optical coupling between two separate condensates. (a)
Bidirectional coupling. Two pairs of Bragg beams with frequencies
$f$ and $f+\nu$ are applied to two separate condensates. (b)
Outcoupling process. Two condensates are trapped at $x=x_1$ and
$x_2$ $(d=x_2-x_1)$. A pair of Bragg beams with frequency
difference $\nu$ generate a continuous atomic beam from each
condensate. The two atomic beams overlap in $x>x_2$, forming a
matter wave interference pattern. $U_1$ ($U_2$) denotes the trap
depth of the left (right) well, $\tilde{\mu_1}$ ($\tilde{\mu_2}$)
the mean-field interaction energy of the left (right) condensate,
and $k_1$ ($k_2$) the wave number of the atomic beam from the left
(right) condensate outside the trap.\label{f:model}}
\end{center}
\end{figure}

\begin{figure}
\begin{center}
\includegraphics{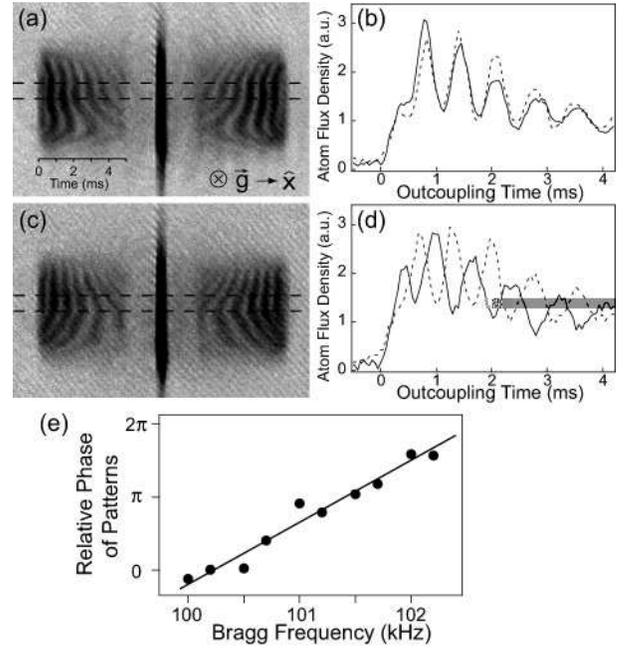}
\caption{Symmetric and antisymmetric correlation between
outcoupled atom patterns. Two pairs of Bragg beams
(Fig.~\ref{f:model}(a)) outcoupled atoms in either $+x$ or $-x$
direction. Absorption images were taken after 5~ms outcoupling
and 2~ms additional ballistic expansion. The left outcoupled atom
patterns were compared with the corresponding right patterns.
Symmetric correlation between two patterns was observed at (a)
$\nu=2\pi\times 100.5$~kHz and (c) antisymmetric at
$\nu=2\pi\times 101.5$~kHz. The field of view is 900~$\mu$m
$\times$ 590~$\mu$m. (b,d) Outcoupled atom flux densities were
obtained by integrating optical densities between the dashed
lines and converting the spatial coordinate to the time
coordinate. The solid (dashed) lines correspond to left (right)
outcoupled atoms. (e) The coupling phase $\theta$ of the two
outcoupled patterns showed a linear dependence on $\nu$ with
$\partial \theta/\partial \nu=(2.4$~kHz$)^{-1}$.\label{f:pattern}}
\end{center}
\end{figure}

In this Letter, we experimentally demonstrate phase-sensitive
optical coupling of two spatially separate Bose-Einstein
condensates, using Bragg scattering. The situation we are
investigating is two condensates, irradiated by two pairs of
Bragg beams (Fig.~\ref{f:model}(a)). The two pairs of Bragg beams
couple out beams of atoms propagating to the left or the right,
respectively, and these unconfined propagating atoms constitute
the intermediate coupling system in our scheme. Depending on the
relative phases of the two condensates and the coupling states,
we observe only one outcoupled beam propagating to one or the
other side, or two identical beams propagating in opposite
directions (Fig.~\ref{f:pattern}). This demonstrates phase
control of currents and establishes a new scheme to realize
Josephson effects with two non-overlapping condensates. In the
following, we present a model for the phase-sensitive outcoupling
process and an experimental test of the prediction that the phase
of the atomic currents into each condensate can be controlled.
Finally, we suggest a Josephson optical coupling of two
condensates in a ring geometry.

First, we elaborate on the situation with a unidirectional
optical coupling (Fig.~\ref{f:model}(b)). We use the conventional
wavefunction description for condensates. Two condensates 1 and 2
are trapped in a double-well potential and optically coupled into
unconfined states by a single pair of Bragg beams. Ignoring the
accumulated phase shifts due to the interaction with the
condensates, we approximate the unconfined coupling states as
truncated free propagating states, i.e., $\psi_i(x,t)\propto
\Theta(x-x_i) \sqrt{\gamma N_i} e^{i\chi_i(x,t)}$ ($i=1,2$),
where $\Theta(x)$ is the Heaviside step function, $\gamma$ is the
outcoupling efficiency of the Bragg beams, $N_i$ is the total
atom number of condensate $i$, and $\chi_i(x,t)=k_i x -\omega_i t
+ \chi_{i0}$ is the phase of the coupling state with
$\hbar\omega_i=\frac{\hbar^2 k_i^2}{2m}$, where $m$ is atomic
mass. The phase continuity at the coupling position $x=x_i$
requires
\begin{equation}\label{e:phase}
    \chi_i(x_i,t) = \phi_i(t)+ \phi_B(x_i,t)-\pi/2,
\end{equation}
where $\phi_i(t)$ is the phase of the condensate, $\phi_B(x,t)=
2k_r x - \nu t + \phi_{B0}$ is the phase of the Bragg beams with
wave number $k_r$ and frequency difference $\nu$, and $-\pi/2$ is
the phase shift attributed to the scattering
process~\cite{VCK03}. In a linear regime with $\gamma \ll 1$,
$\phi_i$ is not perturbed by the coupling, i.e.,
$\phi_i(t)=-\frac{\mu_i }{\hbar}t+\phi_{i0}$, where $\hbar$ is
Planck's constant divided by $2\pi$ and
$\mu_i=-U_i+\tilde{\mu_i}$ (Fig.~\ref{f:model}(b)) is the chemical
potential of the condensate. Satisfying the phase relation
Eq.~(\ref{e:phase}) at all $t$ requires
\begin{eqnarray}
    &&\hbar\omega_i= \hbar\nu + \mu_i,\label{e:timepart}\\
    &&\chi_{i0}=-\delta k_i x_i+\phi_{B0}+\phi_{i0}-\pi/2,
\end{eqnarray}
where $\delta k_i=k_i-2k_r$.  Eq.~(\ref{e:timepart}), the
temporal part in Eq.~(\ref{e:phase}), corresponds to energy
conservation.

In the overlapping region, $x>x_2$, the two atomic beams from
each condensate form a matter wave interference pattern, and the
outcoupled atom density $n(x,t)=|\psi_1(x,t)+\psi_2(x,t)|^2$. For
a better interpretation, we define the right outcoupled atom
density $n_R(s,t)\equiv n(s+x_2,t)$, where $s$ indicates the
distance from the right condensate,
\begin{equation}\label{e:rpattern}
    n_R(s,t)= \frac{\gamma}{2v_r} \big( N_t + 2\sqrt{N_1 N_2}\cos(\Delta k s+\phi_r(t)-\delta k_1 d)\big),
\end{equation}
where $N_t=N_1+N_2$, $\Delta k = k_2 -k_1$, $d=x_2-x_1$, and
$\phi_r(t)=\phi_2(t)-\phi_1(t)$. We approximate the propagating
velocity $v_i=\frac{\hbar k_i}{2m}\simeq 2v_r$ with $\delta k_i
\ll 2k_r$, where $v_r=\frac{\hbar k_r}{m}$ is the recoil
velocity. According to the relative phase $\phi_r$, outcoupled
atoms from the left condensate are coupled into or amplified by
the right condensate. The spatial and temporal modulation of the
outcoupled atom flux $n_R$ represents the evolution of the
relative phase $\phi_r$, which was directly demonstrated in our
previous experiments~\cite{SPS05}.

The phase term $-\delta k_1 d$ can be interpreted as the phase
shift which outcoupled atoms would accumulate during the flight
from the left condensate to the right with respect to the Bragg
beam phase $\phi_B$ which is acting as the phase reference. A
similar relation between $n_R$ and $\phi_r$ can be obtained in
terms of the dynamic structure factor of two separate
condensates~\cite{PS99}, but the phase modulation of coupling
states in the middle of two condensates is likely to be ignored
in the conventional impulse approximation~\cite{ZPS00}. This
phase shift is the key element for an actively-controlled optical
coupling and its physical importance will be manifest in the
following bidirectional coupling scheme.

We now add another pair of Bragg beams to outcouple atoms in the
$-x$ direction. Modifying the above calculation by
$k_{i,r}\rightarrow -k_{i,r}$, the left outcoupled atom density
$n_L(s,t)\equiv n(x_1-s)$ is given as
\begin{equation}
    n_L(s,t)= \frac{\gamma}{2v_r} \big( N_t + 2\sqrt{N_1 N_2}\cos(\Delta k s+\phi_r(t)+\delta k_2 d)\big).
\end{equation}
Considering the atom flux for each condensate, we find rate
equations for $N_1$ and $N_2$. For example, the left condensate
has influx of $\gamma N_2$ from the right condensate and outflux
of $\gamma N_1$ and $n_L(0,t)$ in $+x$ and $-x$ direction,
respectively. The final rate equations read
\begin{equation}
    \dot{N}_{1,2}=-2\gamma\big( N_{1,2} +\sqrt{N_1 N_2}\cos(\phi_r(t)\mp \delta
    k_{1,2}d)\big).
\end{equation}
Except for the global depletion effect of Bragg scattering, the
rate equations describe Josephson oscillations due to the
bidirectional optical coupling, i.e., that the atomic currents
into the condensates depend on the relative phase.

The optical Josephson coupling has a unique feature in the
control of the phase accumulated by atoms in the coupling
state~\cite{IK97}. Since the intermediate system ``delivers" the
phase information from one condensate to the other, the phase can
be manipulated in transit and consequently, the phase of the
effective coupling can be controlled without affecting the two
condensates. In the bidirectional coupling scheme, the control of
the coupling phase is embodied in the phase shift terms, $-\delta
k_1d$ and $\delta k_2d$. We define the coupling phase as $\theta
\equiv(\delta k_1 +\delta k_2)d$, and with $\delta k_i \ll 2k_r$,
approximate $\theta$ as
\begin{equation}
    \theta=\frac{d}{v_r}(\nu -\frac{4E_r}{\hbar}+ \frac{\mu_1+\mu_2}{2\hbar}),
\end{equation}
where $E_r=\frac{\hbar^2 k_r^2}{2m}$ is the recoil energy.
$\theta$ is equivalent to the relative phase of $n_L$ and $n_R$.
When $\theta=0$ ($\theta=\pi$)$\pmod{2\pi}$, $n_L$ and $n_R$ will
show (anti)symmetric correlation.

The control of the coupling phase $\theta$ was experimentally
demonstrated. Condensates of $^{23}$Na atoms in the $|F = 1,m_F =
-1\rangle$ state were prepared in an optical double-well
potential as described in Ref.~\cite{SSP04}. The
$1/e^2$-intensity radius of a focused laser beam for a single
well was $7.6~\mu$m and the typical trap depth was $U_{1,2}\sim
h\times 18$~kHz. The separation of the two wells was
$d=11.4~\mu$m and each well started with a condensate of $\sim
5\times 10^5$ atoms. Two pairs of Bragg beams parallel to the
separation direction were applied to the condensates by
retroreflecting two copropagating laser beams with frequency
difference $\nu$. The lifetime of condensates was over 18~s and
the $1/e$ depletion time due to Bragg scattering into both
directions was 4.5~ms. Outcoupling patterns were measured by
taking absorption images of outcoupled atoms.

When the Bragg frequency difference $\nu$ was varied, the
outcoupling pattern cycled through symmetric and antisymmetric
correlations (Fig.~\ref{f:pattern}). The coupling phase $\theta$
was fit to the observed patterns for each Bragg frequency
(Fig.~\ref{f:pattern}(e)). The linear dependence was measured as
$\partial\theta/\partial\nu=(2.4 \pm 0.2$~kHz$)^{-1}$, which is
consistent with the predicted value $d/v_r=(2.6$~kHz$)^{-1}$.
This clearly demonstrates the presence and control of the
coupling phase in our optical coupling scheme. With the
antisymmetric condition, $\theta=\pi$, as a function of the
propagating relative phase, the output oscillated between
predominantly to the left and predominantly to the right
(Fig.~\ref{f:pattern}(c) and (d)). The experimental situation has
perfect mirror symmetry. Unidirectional output in a symmetric
situation is a macroscopic consequence of the condensates' phase.

Control of the coupling phase can be used to introduce temporal
and spatial variations of Josephson-type coupling. Temporal
control with real-time feedback could ensure the coherent and
continuous replenishment of a condensate (see Ref.~\cite{CSL02}).
For elongated condensates, as used here, spatial control with
barrier heights or well separations could create spatially
varying coupling along the condensate axis, and realize, e.g.,
ring currents.

\begin{figure}
\begin{center}
\includegraphics{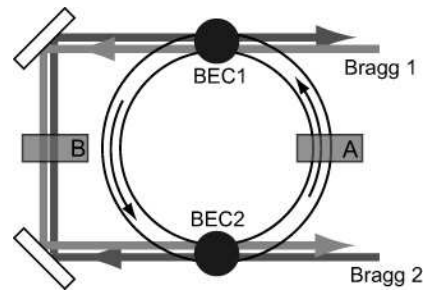}
\caption{Optical coupling of two condensates on a ring. Two
condensates are confined at opposite sides on a ring-shaped
waveguide and a pair of Bragg beams (Bragg 1 and 2) outcouple
atoms in the clockwise direction. The shaded boxes $A$ and $B$
are phase modulators for atoms and photons,
respectively.\label{f:ring}}
\end{center}
\end{figure}

One limitation of the bidirectional coupling scheme is that atoms
are depleted out of the system due to the linear geometry. Even
though the pattern of outcoupled atoms is a crucial signal for
monitoring the coupling dynamics, the coupling time is
fundamentally limited. To overcome this shortcoming, we envisage
a system preserving total atom number like in Fig.~\ref{f:ring},
where atoms circulate between two condensates in a ring
waveguide. With assumptions that the traveling time $\delta t$
for atoms from one condensate to the other is short enough to
satisfy $\dot{\phi_r}\delta t \ll 1$ and that the density
profiles are constant over the trajectories between the two
condensates, the governing equation, in a linear regime, is
\begin{equation}
    \dot{N_2}-\dot{N_1}=2 \gamma \sqrt{N_1 N_2} \cos(\phi_r
    -\phi_m),
\end{equation}
where $\phi_m$ is the effective coupling phase which is
determined by the accumulated phase shift over the round
trajectories and the phase of the Bragg beams.

\begin{figure}
\begin{center}
\includegraphics{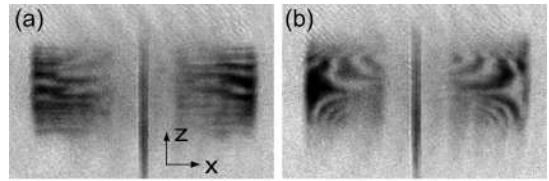}
\caption{Monitoring of slow relative axial motions. Two chemical
potentials were equalized within 400~Hz and the temporal
evolution of the relative phase of two condensates along the
axial $(z)$ direction was recorded in the outcoupled atom
patterns. (a) represents relative dipole oscillation,
corresponding to relative velocity $\approx 300~\mu$m/s or
kinetic energy of $\approx 130$~pK$\times k_B$, and (b) relative
quadruple oscillation. $k_B$ is the Boltzmann
constant.\label{f:lowfreq}}
\end{center}
\end{figure}

The long condensates used here introduce a new degree of freedom
into the usual point-like Josephson junctions: the condensates
can have a spatially varying phase along the axial direction.
Since the optical coupling is selectively established between
condensates at the same axial position, axial gradients of the
relative phase are directly observed through tilted fringes in
the pattern of outcoupled atoms. In Fig.~\ref{f:lowfreq}, we
present two examples showing the effects of relative dipole and
quadruple axial motions of two condensates. Josephson
vortex~\cite{KaKu05} and modulational instabilities~\cite{IsaB05}
in elongated coupled condensates were theoretically suggested.

\begin{figure}
\begin{center}
\includegraphics{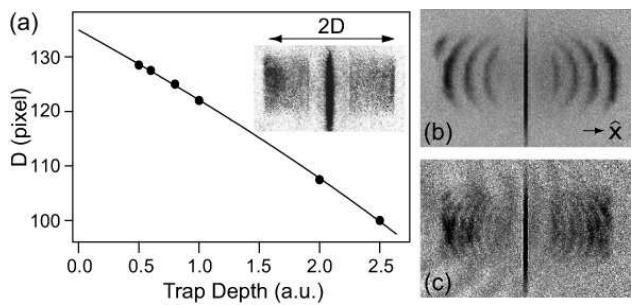}
\caption{Trap characterization by continuous outcoupling. (a)
Atoms were outcoupled from a single well and the traveling
distance $D$ with fixed traveling time $t=7$~ms was measured,
changing the trap depth $U$ by varying the total power of the
laser beam forming the single well. The solid line is a fitting
curve (see the text for details.) with 1~a.u.$=18.0$~kHz and 1
pixel$=3.11~\mu$m. Dipole oscillation in the $x$-direction was
induced by suddenly shifting the trap center. (b) Outcoupling
pattern with Bragg frequency difference $\nu=2\pi\times114$~kHz
and (c) $\nu=2\pi\times101.5$~kHz, which correspond to resonant
velocities of $\approx 4$~mm/s and $\approx 0$~mm/s, respectively.
In (b), the left and the right pattern have antisymmetric
correlation.\label{f:trap}}
\end{center}
\end{figure}

Continuous Bragg scattering was used to characterize the trap
depth and the trap frequency of a single optical trap
(Fig.~\ref{f:trap}). Since momentum and energy imparted in the
scattering process are precisely defined, the kinetic energy of
atoms coupled out of a trap determines the depth of the trap. We
measured the traveling distance $D$ of outcoupled atoms with
fixed traveling time $t$, and determined the trap depth $U$ from
the relation, $D/t=\sqrt{4v_r^2-2U/m}$, ignoring the mean-field
interaction with the condensate and the finite size of the trap.
Additionally, the exact knowledge of the recoil velocity $v_r$
calibrates the optical magnification of images.

On the other hand, the trap frequency was measured using velocity
sensitivity of Bragg scattering. When a condensate oscillates in
a trap, atoms are coupled out only when the condensate is at the
resonant velocity. Since the dipole oscillation of a condensate
in a harmonic trap is the same as the trap frequency $f$, the
outcoupling frequency is the same as $f$ when Bragg beams are
tuned at the maximum velocity (Fig.~\ref{f:trap}(b)), $2f$ at zero
velocity (Fig.~\ref{f:trap}(c)). Even though the frequency
resolution is limited by the finite coupling time, this method
provides a lot of information in a single measurement. For
example, the pattern of outcoupled atoms in Fig.~\ref{f:trap}(b)
is curved because the trap frequency changes along the axial
direction.

The system studied here is perfectly symmetric. Nevertheless, in
any realization of the experiment, the relative phase of the two
condensates assumes a specific value and spontaneously breaks the
symmetry. The unidirectional output is equivalent to the
magnetization in a ferromagnet, which, by spontaneous symmetry
breaking, points into a specific direction. Spontaneous symmetry
breaking can be observed in the interference pattern of two
overlapping condensates which has a definite phase~\cite{ATM97}.
Unidirectional output in a symmetric situation more dramatically
shows the existence of the condensates' phase.

In conclusion, we experimentally studied the optical coupling
between two spatially separate condensates using bidirectional
Bragg scattering and demonstrated that the phase of the coupling
currents can be controlled. This scheme is a new approach for
observing Josephson phenomena, but also for monitoring condensate
motion and characterizing trapping potentials.

This work was funded by NSF, ONR, DARPA, ARO, and NASA. G.-B. J.
acknowledges additional support from the Samsung Lee Kun Hee
Scholarship Foundation, and M.S. from the Swiss National Science
Foundation.

\end{document}